\documentclass[journal=jacsat,manuscript=article]{achemso}

\usepackage[version=3]{mhchem} 
\usepackage{hyperref}
\usepackage{graphicx}
\usepackage{amsmath}
\usepackage{amssymb}
\usepackage{float} 
\usepackage{pdfpages}
\usepackage{subfiles}
\author{Diptabrata Paul}
\email{diptabrata.paul@students.iiserpune.ac.in}
\affiliation[IISER]{Department of Physics, Indian Institute of Science Education and Research (IISER), Pune 411008, India}
\author{Deepak K. Sharma}
\affiliation{Laboratoire Interdisciplinaire Carnot de Bourgogne, UMR 6303 CNRS, Université de Bourgogne Franche-Comté, 9 Avenue Alain Savary, 21000 Dijon, France}
\author{G V Pavan Kumar}
\email{pavan@iiserpune.ac.in}
\affiliation[IISER]{Department of Physics, Indian Institute of Science Education and Research (IISER), Pune 411008, India}
\title{Simultaneous detection of spin and orbital angular momentum of light through scattering from a single silver nanowire}  

\begin{document}

\begin{abstract}

In recent times the spin angular momentum (SAM) and orbital angular momentum (OAM) of light have gained prominence because of their significance in optical communication systems, micromanipulation, sub-wavelength position sensing. To this end, simultaneous detection of SAM and OAM of light beam is one of the important topics of research from both application and fundamental spin-orbit interaction (SOI) point of view. While interferometry and metasurface based approaches have been able to detect the states, our approach involves elastic scattering from a monocrystalline silver nanowire for the simultaneous detection of SAM and OAM state of a circularly polarized Laguerre-Gaussian (LG) beam. By employing Fourier plane (FP) microscopy, the transmitted scattered light intensity distribution in the FP is analyzed to reconstruct the SAM and OAM state unambiguously. The SAM and OAM induced transverse energy flow as well as the polarization dependent scattering characteristics of the nanowire is investigated to understand the detection mechanism. Our method is devoid of complex nanofabrication techniques required for metasurface based approaches and to our knowledge, is a first example of single nano-object based simultaneous SAM and OAM detection. The study will further the understanding of SOI effects and can be useful for on-chip optical detection and manipulation.
\end{abstract}

\maketitle 

\section{Introduction}
A light beam can possess spin angular momentum (SAM) manifested through its circular polarization state, as well as orbital angular momentum (OAM), manifested through its helical phase front \cite{PRA_allen, Molina2007, LPR_Franke}. SAM can have two orthogonal states corresponding to left- and right-circularly polarized light, whereas OAM is unbounded with its infinite orthogonal states, corresponding to the topological charge of vortex beams such as Laguerre-Gaussian beams ($\textrm{LG}^n_m$, $n$ topological charge, $m$ radial index) \cite{Yao_11, Shen2019}. This angular momentum of light corresponds to an energy flow in the beam in the transverse plane with respect to the propagation direction, termed as transverse energy flow \cite{Bekshaev_11}. Interaction of the transverse energy flow of the beam with a medium may lead to spin-orbit interaction (SOI) induced effects \cite{Bliokh2010, Cardano2015, Aiello2015, Bliokh2015} such as directional scattering \cite{Neugebauer2019}, interconversion of angular momentum \cite{Zhao2007, Nechayev2019} as well as micro manipulation \cite{Zhao_09} and position sensing \cite{Oscar2010}. Additionally, beams possessing SAM and OAM have been envisaged for optical communication \cite{Wang2012, Willner_15, Lei2015, Xie2018}, quantum information \cite{Mair2001} applications due to their orthogonal states.\\

To this end, deterministic sensing of SAM and OAM has become one of the important topics of research in recent years. Detection of OAM in macroscopic scale has been achieved through conventional interferometry \cite{Padgett1996,Leach2002,Leach2004} as well as diffraction \cite{Felde_SPIE_2004,Jack2008,deAraujo11,Bekshaev20} and projection through a reciprocal forked diffraction grating \cite{Mair2001}. These methods are constrained by their complicated optical setups. With the advancement of nanofabrication, generation as well as detection of OAM and SAM has been achieved at nanoscale \cite{Mingbo2015, Guo2016, Genevet2012, Chen2018, ZhangLPR2020}. At these length scales, the on chip plasmonic devices offer great advancement from the interferometric based techniques with their lower footprint. But, in most cases these devices involve complex fabrication methods.\\
\begin{figure}[H]
\centering
\includegraphics{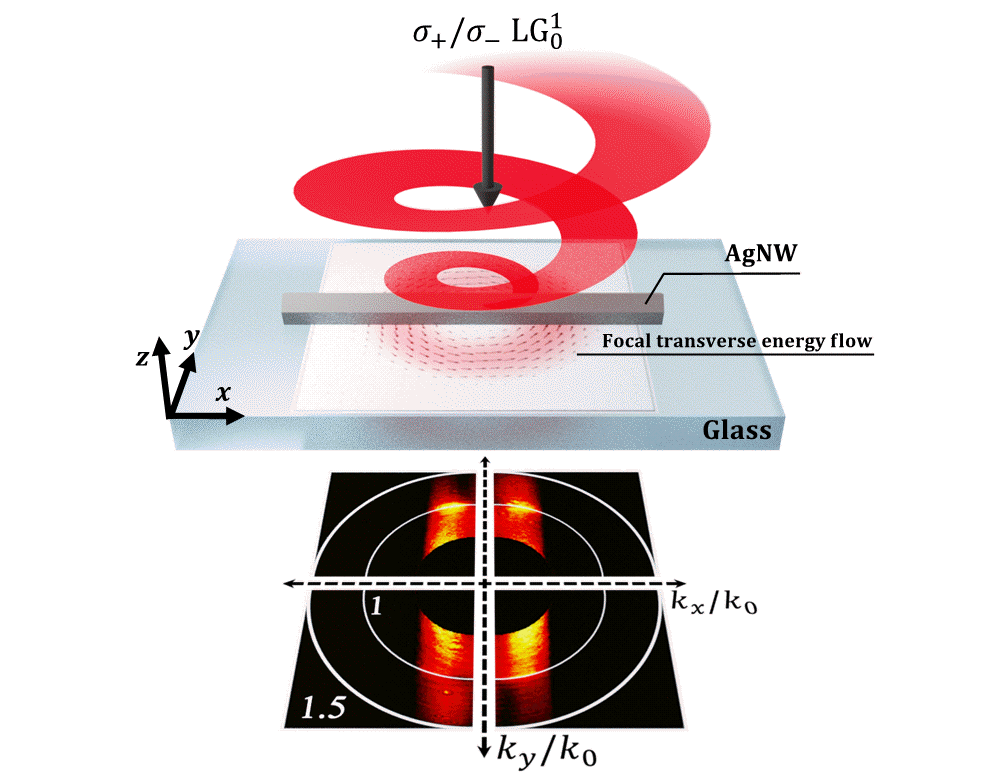}
\caption{Schematic of the experimental configuration. An AgNW placed on a glass substrate scatters the incident LG beam to determine its total angular momentum state.}
\label{f1}
\end{figure}

Thus, there is a requirement of methods which are devoid of these complications and possess smaller footprint for chip-scale applications. Additionally, differential sensing of spin and orbital angular momentum component of a light beam using a single nano-object is yet to be achieved. With this motivation, herein we propose a chemically synthesized quasi-one-dimensional monocrystalline silver nanowire (AgNW) as a sensor for the simultaneous detection of SAM and OAM of light. AgNWs have gained prominence because of their simple preparation method and monocrystalline characteristics. The solution phase synthesis allows them to be self-assembled on any desired substrate and having necessary tunability \cite{Sun2002,Visa2013}. The uniform morphology renders them useful for low loss surface plasmon waveguiding property \cite{Yang2016} and makes them ideal for fundamental photonic circuit applications such as plasmonic waveguide \cite{Yang2016, Chetna2021}, plasmonic antenna \cite{Vasista2018, Tiwari2020}, logic gates \cite{Wei2011} as well as for opto-electrical applications \cite{Lopez2015}. Additionally, in our previous reports we have shown how an AgNW can act as a plasmonic scatterer in sensing total energy flow of an incident beam \cite{Sharma2018,Sharma2019} and as well as probing the inherent spin-Hall effect \cite{Paul2021}.

In this work, a single AgNW dropcasted on a glass substrate has been used as a nanoscopic strip diffractor and the scattered pattern has been extensively analyzed to ascertain the spin and orbital angular momentum state of the incident light, as shown in \textbf{Figure \ref{f1}}. By considering the total transverse energy flow, as well as the individual component due to spin flow and orbital flow at the focus, we are able to distinguish their individual effect on the scattering pattern, allowing us to unambiguously determine their value.

\section{Results and Discussion}

\subsection{Theoretical Calculation}
In order to distinguish spin and orbital angular momentum, we need to characterize their contribution to the energy flow of the beam \cite{Bekshaev_11}. The energy flow of an electromagnetic field, defined by its Poynting vector, can be decomposed into its longitudinal and transverse part. For example, Laguerre-Gaussian (LG) beams, commonly known as vortex beams, possess transverse energy flow corresponding to its orbital flow, owing to its helical wave front. The magnitude and the sense of flow gets dictated by the magnitude and sign of the topological charge of the beam. On the other hand, circularly polarized Gaussian beams exhibit transverse energy flow corresponding to its spin flow, the sense of which gets dictated by the handedness of the circular polarization.\\

\begin{figure*}
\includegraphics{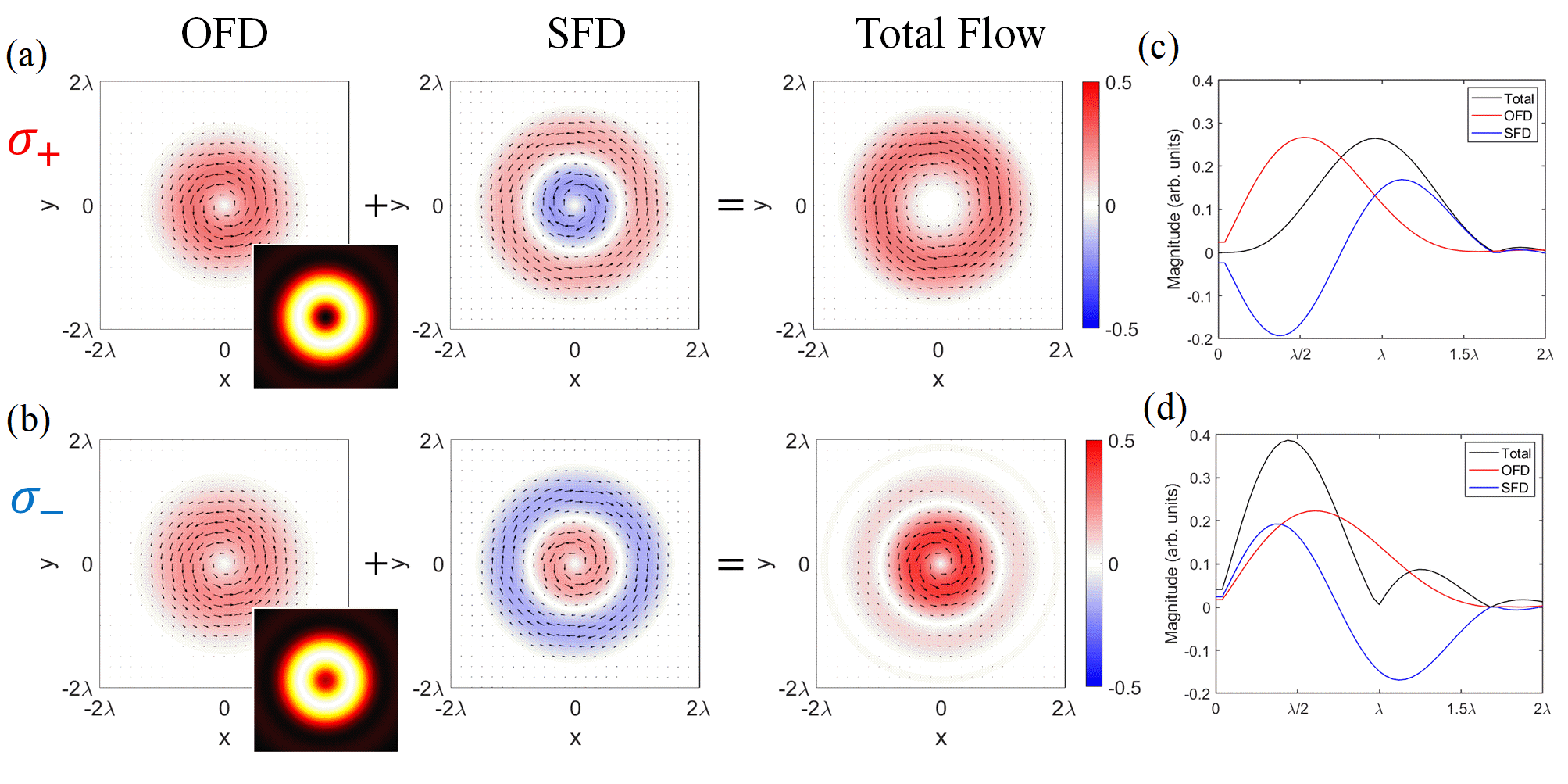}
\caption{Theoretically calculated focal orbital flow density (OFD), spin flow density (SFD) and total transverse energy flow distribution for (a) left-circularly polarized ($\sigma_+$) and, (b) right-circularly polarized ($\sigma_-$) $\textrm{LG}^1_0$ beam. Inset shows the corresponding total field intensity. (c) and (d) represents the OFD, SFD and total transverse energy flow distribution line profile along $x$ axis from $0$ to $2\lambda$.}
\label{f2}
\end{figure*}

To investigate the interaction due to transverse energy flow at nanoscale, we employ the Debye-Wolf integral formulation \cite{Wolf1,WOlf2} to find out the optical electric field ($\mathbf{E}=(E_x,E_y,E_z)$) and magnetic field ($\mathbf{H}= (H_x,H_y,H_z)$) components at the focal plane of a $0.5$ NA lens \cite{Novotny2006}. From these, the corresponding spin flow density (SFD, $\mathbf{{p}^\perp_s}$) and orbital flow density (OFD, $\mathbf{{p}^\perp_o}$) about $z$ axis (propagation direction) can be calculated by considering the transverse component (denoted by superscript $\perp$) of the following equations \cite{Bekshaev_11}:
 
\begin{eqnarray}
\label{eq1}
\mathbf{p_s} = \frac{g}{4\omega}\textrm{Im}[\nabla\times(\varepsilon_0(\mathbf{E^*}\times\mathbf{E}))+\mu_0(\mathbf{H^*}\times\mathbf{H}))],
\end{eqnarray} 

\begin{eqnarray}
\label{eq2}
\mathbf{p_o} = \frac{g}{2\omega}\textrm{Im}[\varepsilon_0((\mathbf{E^*}\cdot(\nabla)\mathbf{E})+\mu_0(\mathbf{H^*}\cdot(\nabla)\mathbf{H})].
\end{eqnarray}

Here $g = (8\pi)^{-1}$ and $\omega$ is the oscillating frequency, $\varepsilon_0$ and $\mu_0$ are the electric permittivity and magnetic permeability of vaccum respectively and $\mathrm{{p}^\perp_{o,i}}\propto \textrm{Im}[\Sigma_j{E}^*_j(\nabla_i) {E}_j]$ ($i,j = (x,y,z)$). \textbf{Figure \ref{f2}(a)} and \textbf{(b)} show the calculated OFD, SFD and total transverse energy flow density ($\mathbf{p}^\perp = \mathbf{p_s}^\perp+\mathbf{p_o}^\perp$) corresponding to left-circularly polarized ($\sigma_+$) and right-circularly polarized ($\sigma_-$) $\textrm{LG}^1_0$ beams at wavelength $\lambda = 633$ nm. The total electric field intensity at the focal plane is shown in the inset. As expected, for both $\sigma_+$ and $\sigma_-$ polarized $\textrm{LG}^1_0$ beam, the OFD has similar magnitude and similar counter-clock orbital flow sense. But, for $\sigma_+$ $\textrm{LG}^1_0$ beam, the SFD and the corresponding spin flow have both clockwise and counter-clock energy flow in its inner and outer part respectively. Additionally, the magnitude of SFD is higher in its inner radius ($|\mathbf{p_s}^\perp| = 0.1085$ at $R = 0.322\lambda$) and slightly lower in its outer radius ($|\mathbf{p_s}^\perp| = 0.0731$ at $R = 1.068\lambda$). It is exactly opposite for the $\sigma_-$ $\textrm{LG}^1_0$ beam. Hence, for $\sigma_+$ $\textrm{LG}^1_0$ beam, counter-clock orbital flow towards the center cancels the clockwise spin flow, whereas, in the outer part the counter-clock spin flow gets added up with orbital flow, resulting an overall counter clock total transverse energy flow having a maximum magnitude of $|\mathbf{p}^\perp| = 0.1515$ at radius $R = 0.868 \lambda$. For $\sigma_-$ $\textrm{LG}^1_0$ the counter-clock inner spin flow gets added up with the counter-clock orbital flow resulting in higher counter-clock transverse energy flow, $|\mathbf{p}^\perp|=0.2551$ at $R = 0.4\lambda$. The OFD, SFD and the total transverse energy flow density radial line profile at the focal plane from $0$ to $2\lambda$ corresponding to $\sigma_+$ and $\sigma_-$ polarized $\textrm{LG}^1_0$ beams are shown in \textbf{figure \ref{f2}(c)} and \textbf{(d)} respectively. In this context, it is important to point out the subtle differences of the intensity profile for $\sigma_+$ and $\sigma_-$ polarized $\textrm{LG}^1_0$ beams as focusing of circularly polarized LG beam aids spin-to-orbital angular momentum transformation \cite{Zhao2007}. The field calculations and corresponding phase profiles are shown in the supplementary information S1.\\

\subsection{Experimental Implementation}
\begin{figure}[t]
\centering
\includegraphics[width=420pt]{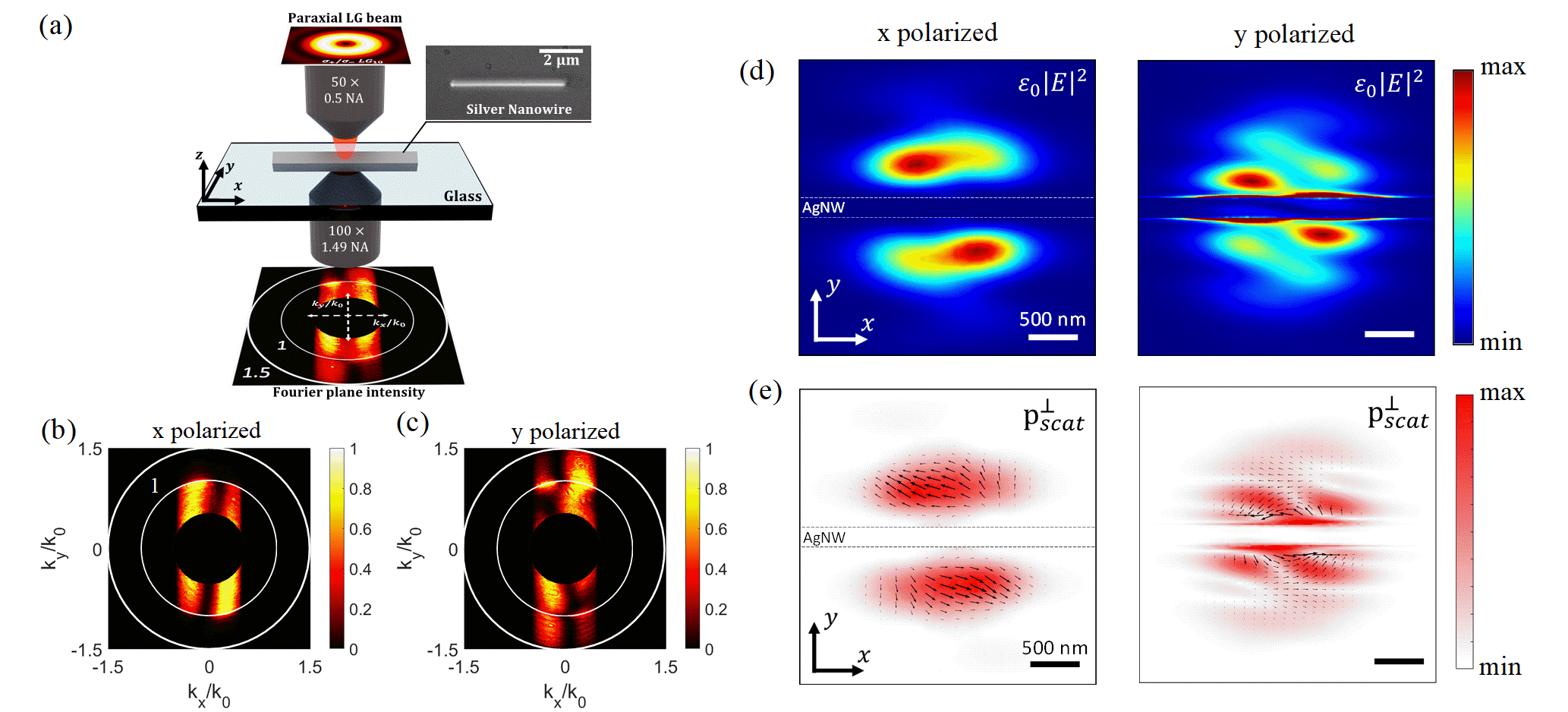}
\caption{(a) Experimental configuration of measurement scheme for the Fourier plane (FP). Inset shows the bright field optical micrograph of AgNW. Experimentally measured Fourier plane intensity distribution for scattering of (b) $x$ polarized and, (c) $y$ polarized $\textrm{LG}^1_0$ beam. Inner and outer white circles represent the critical angle at the air-glass interface and the collection limit of the objective lens NA respectively. Incident NA region is blocked and appears as a black disk at the center. (d) Simulated near-field electric field intensity distribution at the air-glass interface for $x$ and $y$ polarized incident beam. (e) Corresponding scattered total transverse energy flow density ($\textrm{p}^{\perp}_{scat}$) distribution for $x$ and $y$ polarized $\textrm{LG}^1_0$ beam. The black arrows indicate the energy flow direction.}
\label{f3}
\end{figure}

The focal optical field and its orbital and spin flow constituents, upon interaction with a nano structure will lead to flow dependant scattering allowing us to determine its state. Although different nano-structures have been shown to have exhibited transverse energy flow dependant scattering \cite{Genevet2012,Kerber2017, Chen2018, Sharma2018, Sharma2019}, differential sensing of spin and orbital flow is yet to be achieved by a single nano-object. To that end, we use a single crystalline silver nanowire (AgNW) to scatter the incident optical field and sense its properties. The implementation of forward scattering configuration is achieved by dropcasting AgNWs of width $\approx 350$ nm onto a glass cover-slip, which acts as a nanoscopic strip scatterer \cite{Felde_SPIE_2004,Bekshaev20} for the incident optical fields. A $50\times 0.5$ NA objective lens has been used to focus the incident beam and the scattered signal was collected using a $100\times 1.49$ NA lens, in the transmission configuration (\textbf{Figure \ref{f3}(a)}). A low power of $0.13 \mathrm{\mu W}$ has been used for the scattering of $\textrm{LG}^{\pm 1}_0$ beam. Fourier plane (FP) intensity distribution of the forward scattered light was collected by projecting the back focal plane of the collection objective lens onto a CCD using relay optics \cite{Kurvits_15, FP_adarsh2019}. The collected FP intensity distribution can be divided into two regions, sub-critical region ($\textrm{NA}<1$) and super-critical region ($\textrm{NA}>1$). The incident NA part (NA $0-0.5$) in the sub-critical region is dominated by the un-scattered incident light, and is omitted from our analysis, appears as a black disk at the center of the measured FP. Details of the experimental configuration is elaborated in the experimental section and supplementary information S2.\\

\subsection{Polarization dependence of scattering intensity}
To determine the contribution due SAM and OAM in the scattered FP, it is imperative to understand the scattered FP patterns of its constituent linear poalrization components. The longitudinally polarized optical field ($E_x$) with respect to the nanowire long axis ($x$ axis) gets scattered in the sub-critical region, as shown in the experimentally measured data in \textbf{figure \ref{f3}(b)} for $x$ polarized $\textrm{LG}^1_0$ beam. On the other hand, the transversely polarized fields ($E_y$) lead to near-field (NF) at the NW edges, and consequent signal at the super-critical region in the FP as well as sub-critical region, shown in \textbf{figure \ref{f3}(c)} for $y$ polarized $\textrm{LG}^1_0$ beam. Additionally, the polarization of the scattered light for the $x$ polarized incident beam is predominantly $x$ polarized and similarly, for $y$ polarized beam scattering, the light scattered in the sub- and super-critical region is mostly $y$ polarized (supplementary information S3). Thus, for scattering of circularly polarized $\textrm{LG}^1_0$ beam, the state of circular polarization can be determined by analyzing the sub-critical region intensity distribution as only in this regime both in-plane $x$ and $y$ components of polarization scatter. Whereas, in super-critical region, predominantly scattered light is from $y$ polarized field and will posses information about OAM state of the incident beam. This has been further confirmed by analyzing the FP due to $\sigma_+$ polarized $\textrm{LG}^1_0$ beam (supplementary information S3), where upon analyzing the FP intensity distribution for $x$ and $y$ polarized components, we can recover the FP intensity distribution for $x$ and $y$ polarized intensity distribution as shown in figure \ref{f3}(b) and (c) respectively. It must also be observed that both FP intensity distribution for $x$ and $y$ polarized $\textrm{LG}^1_0$ beam exhibit two lobes as well as preferential scattering pattern due to their orbital flow \cite{Sharma2019}. But, the preference for $x$ polarized case is opposite to that of $y$ polarized case. This can be attributed to the polarization dependent interaction of the NW with the incident beam as shown through numerical simulation in \textbf{figure \ref{f3}(d)} and \textbf{(e)}. The details of the finite element method numerical simulation and geometry is given in supplementary information S4. For $x$ polarized $\textrm{LG}^1_0$ beam the field gets scattered from the NW and the coresponding total transverse energy flow ($\textrm{p}^{\perp}_{scat}$) points away from the NW and opposite to the incident beam $\textrm{p}^{\perp}$. On the other hand, $y$ polarized $\textrm{LG}^1_0$ beam aids the interaction of the transverse energy flow with the nanowire and leads to generation of evanescent near-field at nanowire edges, apparent with the presence of scattered light signal in the super-critical region of FP in figure \ref{f3}(c). Consequently, the scattered field $\textrm{p}^{\perp}_{scat}$ flow is along the same direction as that of the incident beam. Thus, a nanowire through its scattering can detect OAM modes as well as its linear polarization states. The effect is qualitatively similar to a nanoscopic strip scatterer \cite{Felde_SPIE_2004,Bekshaev20}.

\begin{figure*}[t]
\centering
\includegraphics[width =350pt ]{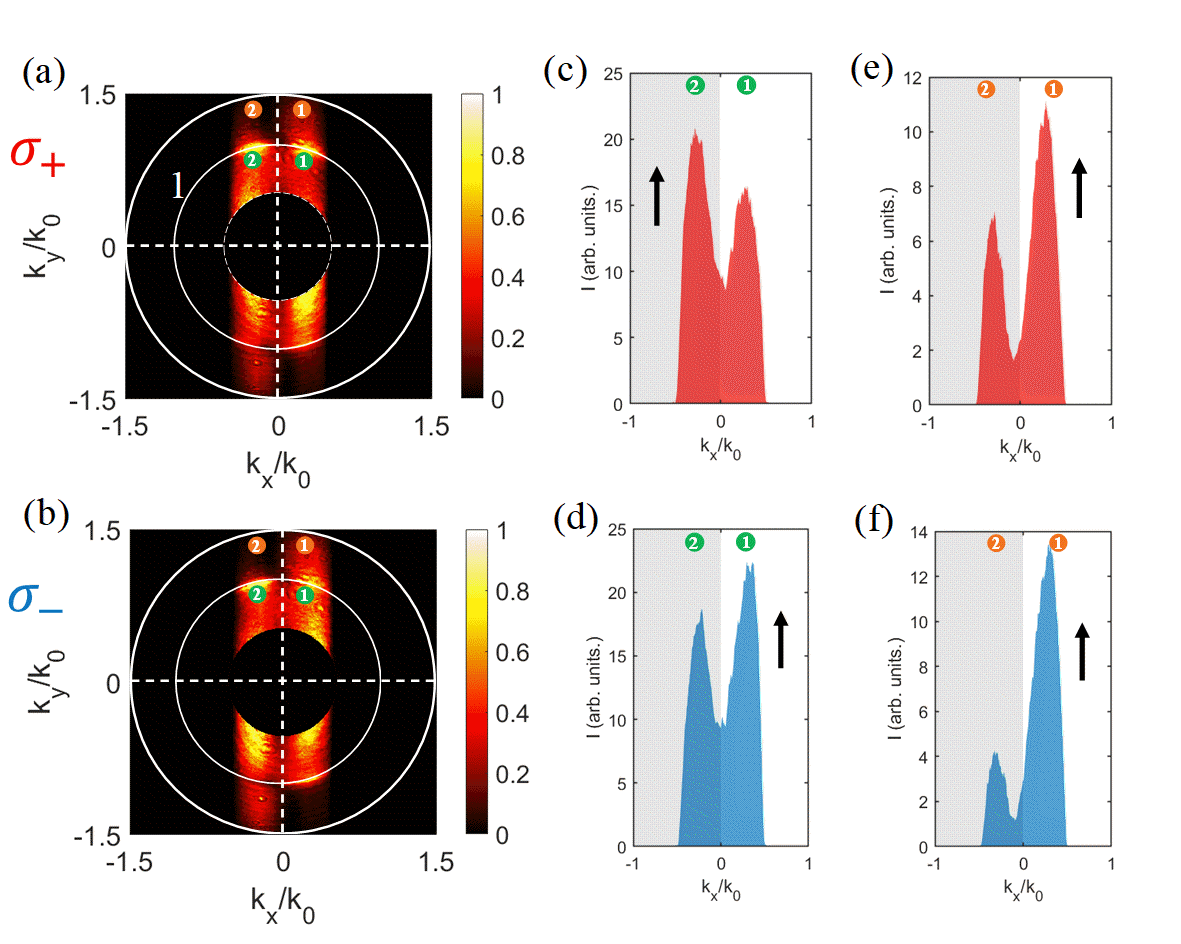}
\caption{Experimentally measured scattered FP intensity distribution of (a) $\sigma_+$ polarized and (b) $\sigma_-$ polarized $\textrm{LG}^1_0$ beam. The corresponding intensity distribution profile in the sub-critical region of the $k_y>0$ part of the FP for $\sigma_+$ and $\sigma_-$ polarized $\textrm{LG}^1_0$ beam is shown in (c) and (d) respectively. The same for super-critical region is shown in (e) and (f). The black arrows indicate the effective directionality.}
\label{f4}
\end{figure*}

\subsection{Nanowire as a sensor for OAM and SAM state }
Next, we investigate the FP intensity distribution for $\sigma_+$ and $\sigma_-$ polarized $\textrm{LG}^{\pm1}_0$ beam to determine OAM and SAM state of the beam. \textbf{Figure \ref{f4}(a)} and \textbf{(b)} shows the FP intensity distribution for $\sigma_+$ and $\sigma_-$ $\textrm{LG}^1_0$ beam respectively. We analyze the pattern by dividing the FP into four parts and looking into the intensity distribution of either upper ($k_y>0$ region) or lower half ($k_y<0$ region). Additionally, the SAM and OAM signature can be obtained by analysing the signal in the sub-critical and super-critical region intensity distribution respectively. \textbf{Figure \ref{f4}(c)} and \textbf{(d)} shows the intensity distribution profile within the sub-critical region of the $k_y>0$ part in the FP, for $\sigma_+$ and $\sigma_-$ polarized $\textrm{LG}^1_0$ beam respectively. The preference of the scattering is opposite for $\sigma_+$ polarized $\textrm{LG}^1_0$ beam scattering with respect to that of $\sigma_-$ $\textrm{LG}^1_0$ beam data as indicated by the black arrows. The preference is the result of spin flow. Since the dominant contribution tot he spin flow occurs  due to the inner circular part of the beam  as shown in figure 2, the preference follows the same handedness. The super-critical region intensity distribution profile in the $k_y>0$ half for both $\sigma_+$ and $\sigma_-$ polarized $\textrm{LG}^1_0$ beam exhibit similar preference as shown in \textbf{Figure \ref{f4} (e) and (f)}. This preference as well as the number of lobes indicates the OAM state of the incident beam \cite{Sharma2019}.

\begin{figure*}[t]
\centering
\includegraphics[width =350pt ]{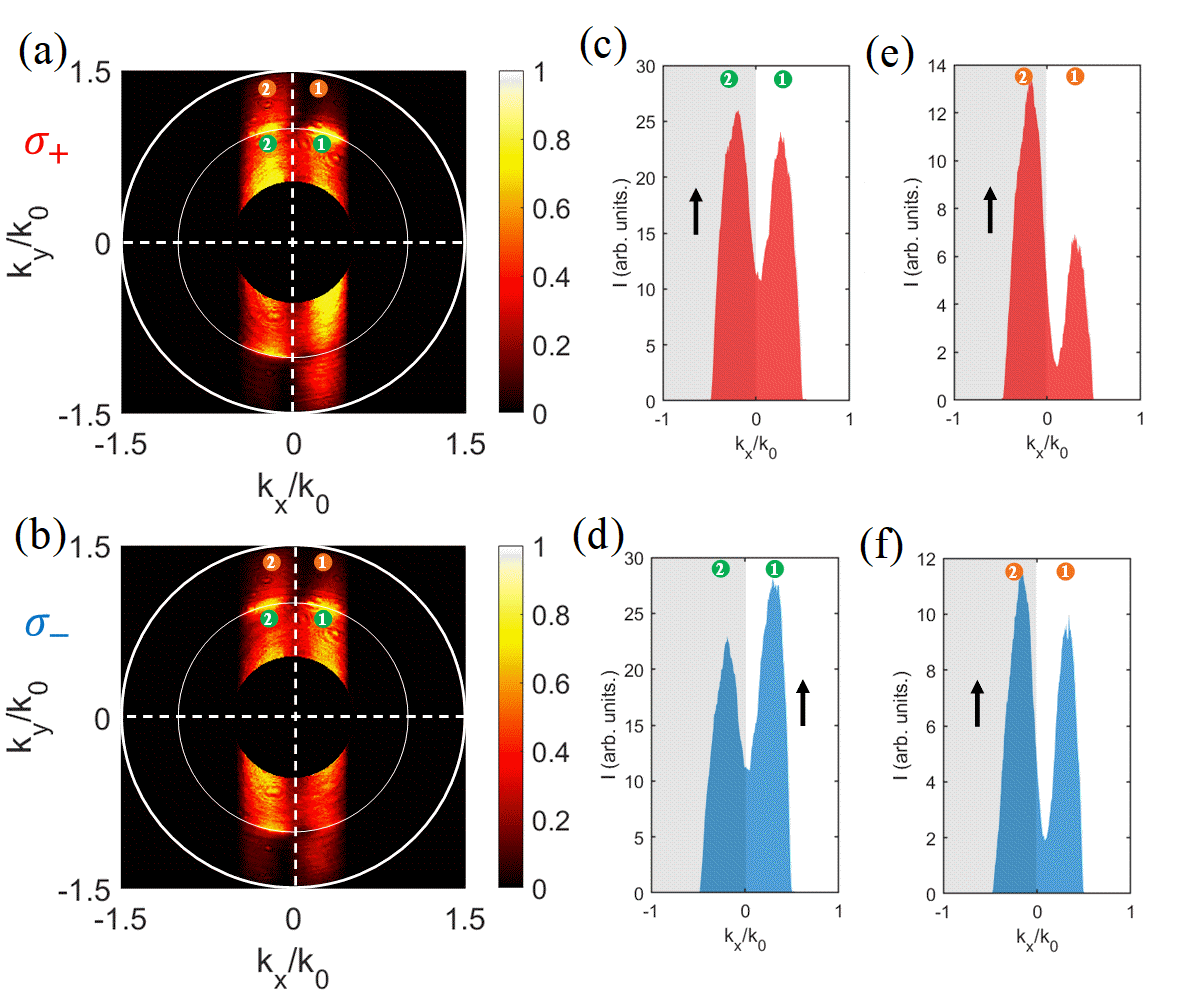}
\caption{Experimentally measured scattered FP intensity distribution (a) $\sigma_+$ and (b) $\sigma_-$ $\textrm{LG}^{-1}_0$ beam.  The corresponding intensity distribution profile in the sub-critical region of the $k_y>0$ part of the FP for $\sigma_+$ and $\sigma_-$ polarized $\textrm{LG}^{-1}_0$ beam is shown in (c) and (d) respectively. The same for super-critical region is shown in (e) and (f). The black arrows indicate the effective directionality.}
\label{f5}
\end{figure*}

The intensity distributions can be further quantified by calculating the directionality, $D=(I_{2}-I_{1})/(I_{2}+I_{1})$, where $I_{i}$ represents the integrated intensity of $i^{th}$ lobe for sub- or super-critical region. From the experimentally measured data in figure \ref{f4} (c) and (d) we find, $D_{sub}=0.094$ for $\sigma_+$ $\textrm{LG}^1_0$ beam and $D_{sub}=-0.123$ for $\sigma_-$ $\textrm{LG}^1_0$ beam in the sub-critical region. Thus, $D_{sub}>0$ for $\sigma_+$ polarization case and $D_{sub}<0$ for $\sigma_-$ polarized beam, i.e., sign of the directionality in the sub-critical region inverts depending upon the spin flow direction and hence the SAM state. On the other hand, the preferential scattering in the super-critical region of the FP indicates the sense of orbital flow and hence the OAM state: $D_{super} = -0.284$ for $\sigma_+$ polarized and $D_{super} = -0.563$ for $\sigma_-$ polarized $\textrm{LG}^1_0$ beam, i.e., directionality sign ($D_{super}<0$) remains the same. Hence, $D_{sub}>0$ and $D_{super}<0$ indicates $\sigma_+$ $\textrm{LG}^1_0$ beam excitation and $D_{sub}<0$ and $D_{super}<0$ indicates $\sigma_-$ $\textrm{LG}^1_0$ beam excitation. Although the differential sensing of the orbital and spin flow is achieved, the total transverse energy flow influences the effective directionality magnitude in the super-critical region to some extent. The measured directionality for the super-critical region is significantly lower for $\sigma_+$ $\textrm{LG}^1_0$ beam than for $\sigma_-$ $\textrm{LG}^1_0$ beam, which is due to the higher total transverse energy flow density for $\sigma_-$ polarized beam, as shown in figure \ref{f2}. For sub-critical region the change is much smaller due lower directionality value as well as positional sensitivity. The directionality values will be exactly opposite in sign when we consider the lower half ($k_y<0$ region) of the scattered FP intensity distribution as the intensity pattern for $k_y>0$ region is anti-symmetric with respect to $k_y<0$ region (Supplementary information S5). 

The detection mechanism is robust with respect to the inversion of the orbital flow. We prove this by investigating the scattered FP intensity distribution for $\sigma_+$ polarized and $\sigma_-$ polarized $\textrm{LG}^{-1}_0$ beam as shown in \textbf{Figure \ref{f5}(a)} and \textbf{(b)} respectively. The corresponding intensity distribution profile in sub critical region of the the $k_y>0$ half for the $\sigma_+$ and $\sigma_-$ polarized case is shown in \textbf{Figure \ref{f5} (c)} and \textbf{(d)} respectively. The measured directionality is found to be $D_{sub} = 0.054$ for $\sigma_+$ and $D_{sub} = -0.132$ for $\sigma_-$ polarized $\textrm{LG}^{-1}_0$ beam. The preference of scattering in this region is same as that of the $\sigma_+$ and $\sigma_-$ polarized $\textrm{LG}^1_0$ beam, indicating the same SAM sign. Whereas, for the super-critical region, the intensity distribution profile shown in \textbf{Figure \ref{f5} (e)} ($D_{super} = 0.369$) and \textbf{(f)} ($D_{super} = 0.114$), the preference is opposite to that of the $\textrm{LG}^1_0$ case, indicating the sign inversion of the topological charge  and hence the orbital as well as total transverse energy flow of the beam. Therefore, owing to the polarization and transverse energy flow dependent interaction of the AgNW with the incident beam, we are able to unambiguously detect OAM and SAM of the incident beam. The accuracy of the simultaneous detection of the SAM and OAM state depends upon the width of the nanowire used for the scattering process. We have performed the experiments for nanowires having width $\approx 200$ nm, and the results have been elaborated in supplementary information S6. In general, the directionality in the sub- and super-critical region depends upon the incident polarization state of the beam. For $\textrm{LG}^1_0$ beam, the elliptical polarization dependence of the directionality and the reconstruction of elliptical polarization states are investigated in supplementary information S7.
\\

\begin{figure*}[t]
\centering
\includegraphics[width =350pt ]{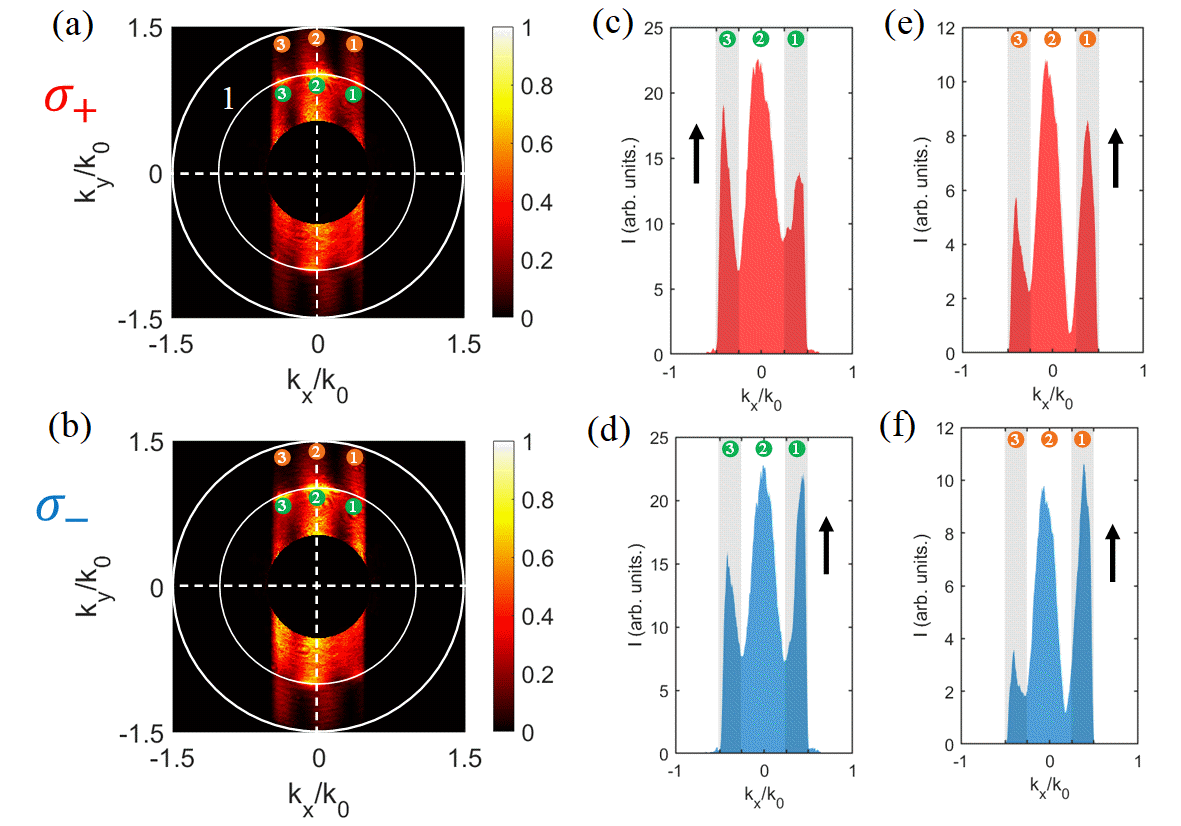}
\caption{Experimentally measured scattered FP intensity distribution of (a) $\sigma_+$ polarized and (b) $\sigma_-$ polarized $\textrm{LG}^2_0$ beam. The corresponding intensity distribution profile in the sub-critical region of the $k_y>0$ part of the FP for $\sigma_+$ and $\sigma_-$ polarized $\textrm{LG}^2_0$ beam is shown in (c) and (d) respectively. The same for super-critical region is shown in (e) and (f). The black arrows indicate the effective directionality.}
\label{f6}
\end{figure*}

\subsection{Detection of OAM and SAM for higher order LG beam}
Next, we explore the possibility of  simultaneous detection of higher order OAM as well as SAM and investigate the FP intensity distribution for higher order LG beam: $\textrm{LG}^2_0$ beam. \textbf{Figure \ref{f6}(a)} and \textbf{(b)} shows the FP intensity distribution for $\sigma_+$ and $\sigma_-$ polarized $\textrm{LG}^2_0$ beam respectively. Here we observe 3 lobes, instead of 2 like in the case of $\textrm{LG}^1_0$ beam, indicating the magnitude of the topological charge ($n+1$ lobes for $\textrm{LG}^n_0$ beam)\cite{Sharma2019}. The corresponding intensity distribution in the sub-critical region of the $k_y>0$ half is shown in \textbf{Figure \ref{f6} (c)} and \textbf{(d)}. The same for the super-critical region is shown in \textbf{figure \ref{f6}(e)} and \textbf{(f)}. From both figure 6(c) and (d) it can be seen that the central lobe remains of almost same intensity for $\sigma_+$ and $\sigma_-$ polarized case, whereas the gray-shaded right-most (numbered $1$) and left-most (numbered $3$) lobes undergo intensity change. Thus we consider the integrated intensity of these two lobes only (1: NA $0.5$ to $0.25$ and 2: NA $-0.25$ to $-0.5$). The directionality values for the sub-critical region intensity distribution in figure \ref{f6} (c) and (d) are $D_{sub} = 0.029$ and $D_{sub} = -0.157$ respectively. As before, the sign inversion of directionality allows us to accurately detect the SAM sign and the signs are same as in the case of $\textrm{LG}^1_0$ beam. On the other hand, the directionality for the super-critical region, $D_{super} = -0.278$ for $\sigma_+$ polarized and $D_{super} = -0.555$ for $\sigma_-$ polarized case indicates the OAM sign.\\

It is important to notice that the directionality values we get in the sub-critical region for $\sigma_+$ polarized beam is lower for $\textrm{LG}^2_0$ beam than that of $\textrm{LG}^1_0$ beam. This is because, as the topological charge magnitude of LG beam increases, the magnitude of the orbital flow increases (supplementary information S8), and dominates the total energy flow and hence it will be more difficult to distinguish the effect of the spin flow. Additionally, presence of $n+1$ lobes  for large $n$ for $\textrm{LG}^n_0$ beam scattering \cite{Sharma2019} in a small scattered width ($k_x =$ $-0.5$ to $0.5$) in the FP intensity distribution, will make it harder for us to detect the fluctuation in the polarization dependent magnitude of the lobes.

\section{Conclusion}

To summarize, we have experimentally shown how forward scattering from a single AgNW can be used for simultaneous unambiguous sensing of spin and orbital angular momentum by employing Fourier plane microscopy. By analyzing the intensity lobes due to circularly polarized LG beam in the sub- and super-ciritical region in the Fourier plane, we show that the preference of scattering in the sub-critical region changes when the spin is inverted, whereas the super-critical region intensity distribution preference depends predominantly on the orbital energy flow of the incident beam, thus allowing us unambiguous detection of both. The incident linear polarization dependent behaviour of the AgNW of scattering within the sub-critical region or both sub- and super-critical region in the FP, allows the detection mechanism. The focal optical field calculation gives us more insight onto the scattering mechanism as well as the resulting directionality. Thus, our work emphasizes spin-orbit interaction based simple single element chip-scale detection configuration with prominent spin and orbital angular momentum detection capability. Additionally, the spin-orbit interaction dependant scattering may lead to opportunities in harnessing the resulting optical forces for optical manipulation at nano scale as well as furthering our understanding of spin-orbit interaction.

\section*{Acknowledgements}
This work was partially funded by Air Force Research Laboratory grant (FA2386-18-1-4118 R\&D 18IOA118), DST Energy Science grant (SR/NM/TP-13/2016) and Swarnajayanti fellowship grant (DST/SJF/PSA-02/2017-18). DP and DKS acknowledges Sunny Tiwari, Rahul Chand, Vandana Sharma, Shailendra K. Chaubey and Chetna Taneja for fruitful discussions.

\bibliography{ref1.bib}
\newpage


\section*{Supplementary material}

# Supplementary information: Simultaneous detection of spin and orbital angular momentum of light through scattering from a single silver nanowire

Diptabrata Paul,[1] Deepak K Sharma,[2] and G V Pavan Kumar[1]

[1]*Department of Physics, Indian Institute of Science Education and Research (IISER), Pune 411008, India*
[2]*Laboratoire Interdisciplinaire Carnot de Bourgogne, UMR 6303 CNRS,*
*Université de Bourgogne Franche-Comté, 9 Avenue Alain Savary, 21000 Dijon, France*

## Contents

## S1: FIELD CALCULATION FOR FOCUSED BEAMS

The fields are calculated employing Debye-Wolf integral representation [1, 2] and closely follows [3]. When an paraxial light beam with electric field $\mathbf{E_{in}}$ and wave vector $\mathbf{k} = k\hat{\mathbf{z}}$ passing through medium of refractive index $n_1$ is incident on an objective lens of numerical aperture, $\mathrm{NA} = n_1 \sin(\theta_{\max})$ and having focal length $f$, then the complex field distribution at the focal plane can be obtained by,

$$\mathbf{E}(\rho, \varphi, z) = -\frac{\mathrm{i}kf\mathrm{e}^{-\mathrm{i}kf}}{2\pi} \int\limits_{0}^{\theta_{\max}} \int\limits_{0}^{2\pi} \mathbf{E_{ref}}(\theta, \phi) \mathrm{e}^{\mathrm{i}kz\cos\theta} \mathrm{e}^{\mathrm{i}k\rho\sin\theta\cos(\phi-\varphi)} \sin\theta \, \mathrm{d}\phi \, \mathrm{d}\theta \tag{1a}$$

$$\mathbf{E_{ref}} = [t^{\mathrm{s}}[\mathbf{E_{in}} \cdot \mathbf{n}_\phi]\mathbf{n}_\phi + t^{\mathrm{p}}[\mathbf{E_{in}} \cdot \mathbf{n}_\rho]\mathbf{n}_\theta]\sqrt{\cos\theta} \tag{1b}$$

The formulation takes into account both the refraction of the paraxial beam at the spherical lens surface (reference sphere) given by $\mathbf{E_{ref}}(\theta, \phi)$ and the non-paraxial effects there after. $t^{\mathrm{s}}$ and $t^{\mathrm{p}}$ are the transmission coefficients corresponding to s and p polarized electric fields respectively. $\mathbf{n}_\rho$ and $\mathbf{n}_\phi$ represents the unit vectors of a cylindrical coordinate system whereas the unit vectors of a spherical polar coordinate are given by $\mathbf{n}_\theta$ and $\mathbf{n}_\phi$.

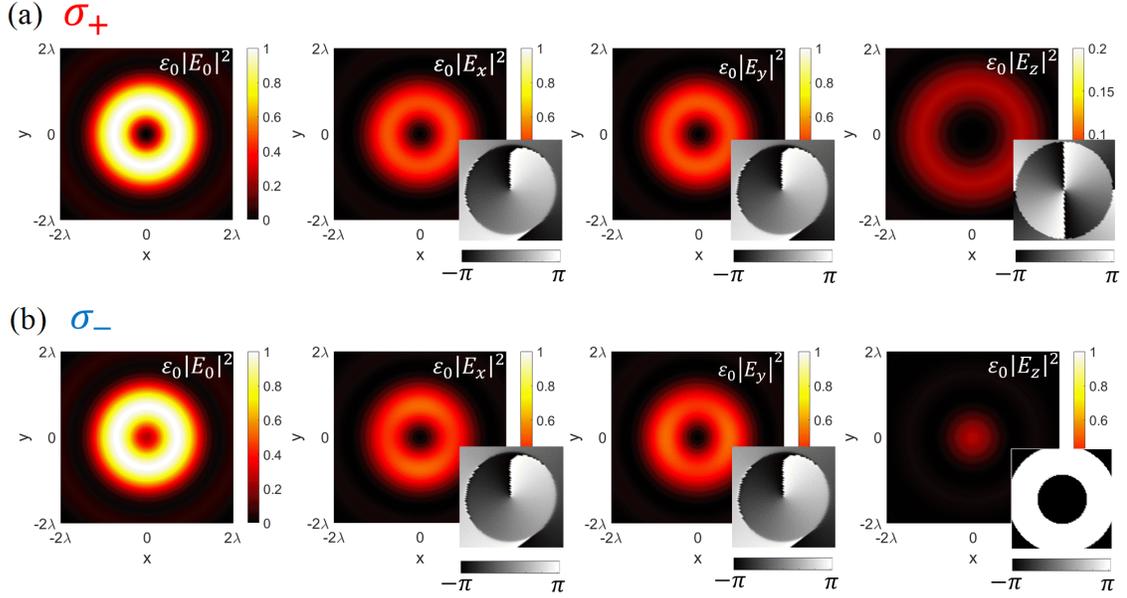

FIG. S1: Theoretically calculated focal electric field intensity distributions and corresponding phase profiles of (a) left-circularly polarized ($\sigma_+$) and (b) right-circularly polarized ($\sigma_-$) $\mathrm{LG}_0^1$ beam. The $E_z$ field corresponding to $\sigma_+$ polarization has phase singularity where as the $\sigma_-$ $E_z$ doesn't possess phase singularity, indicating spin-to-orbit conversion.

## S2: EXPERIMENTAL CONFIGURATION

The forward scattering measurements is performed using a home-built dual-channel Fourier microscope. A $50\times$ 0.5 NA objective lens (Olympus) was used to focus the incident light on a silver nanowire drop-casted glass coverslip placed on a piezo stage (PI P-733.3CD XY Piezo Nanopositioner). The scattered light has been collected using $100\times$ 1.49 NA lens and projected to the CCD (Thorlabs) using relay optics. A spatial light modulator (Holoeye LC-R 2500 SLM) with off axis blazed hologram has been used to modulate the phase of the 632.8 nm Gaussian beam to produce LG beams of different order ($\mathrm{LG}_0^n$, topological charge $n$). The first order diffracted beam was filtered through a pin hole and routed towards the microscope objective lens.

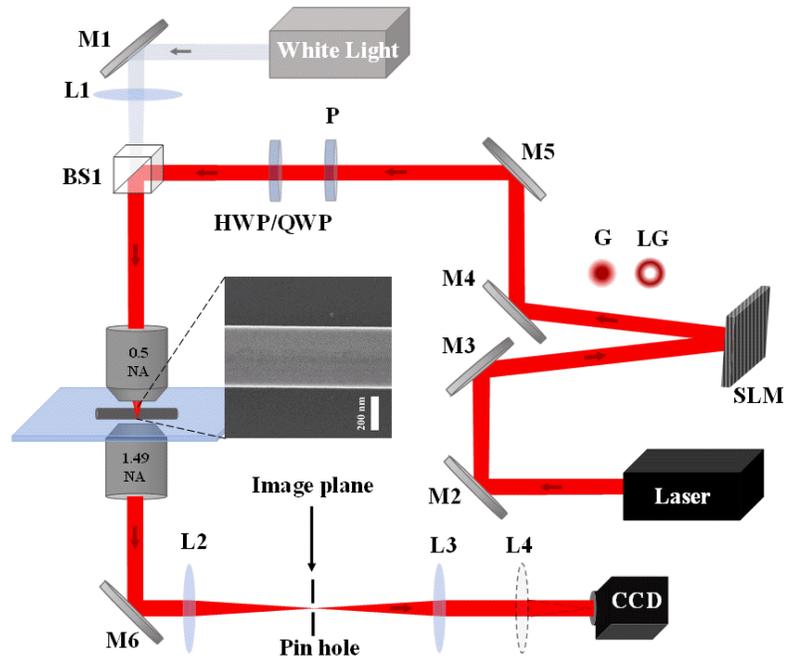

FIG. S2: Detailed optical setup for forward-scattering measurements. M1-M6 are mirros; HWP: Half-wavep plate; QWP: Quarter-wave plate; BS1: Beam-splitter. A $50 \times 0.5$ NA objective lens focuses the incident Gaussian/$LG_0^{\pm 1}$ beam produced through spatial light modulator(SLM) onto the sample, placed on a piezo stage. The scattered light is collected through $100 \times 1.49$ NA objective lens and the back-focal plane projected onto the CCD using relay optics for Fourier plane imaging. The setup is same as used in our previous reports [4, 5]

## S3: ANALYZED FOURIER PLANE FOR LG$_0^1$ EXCITATION

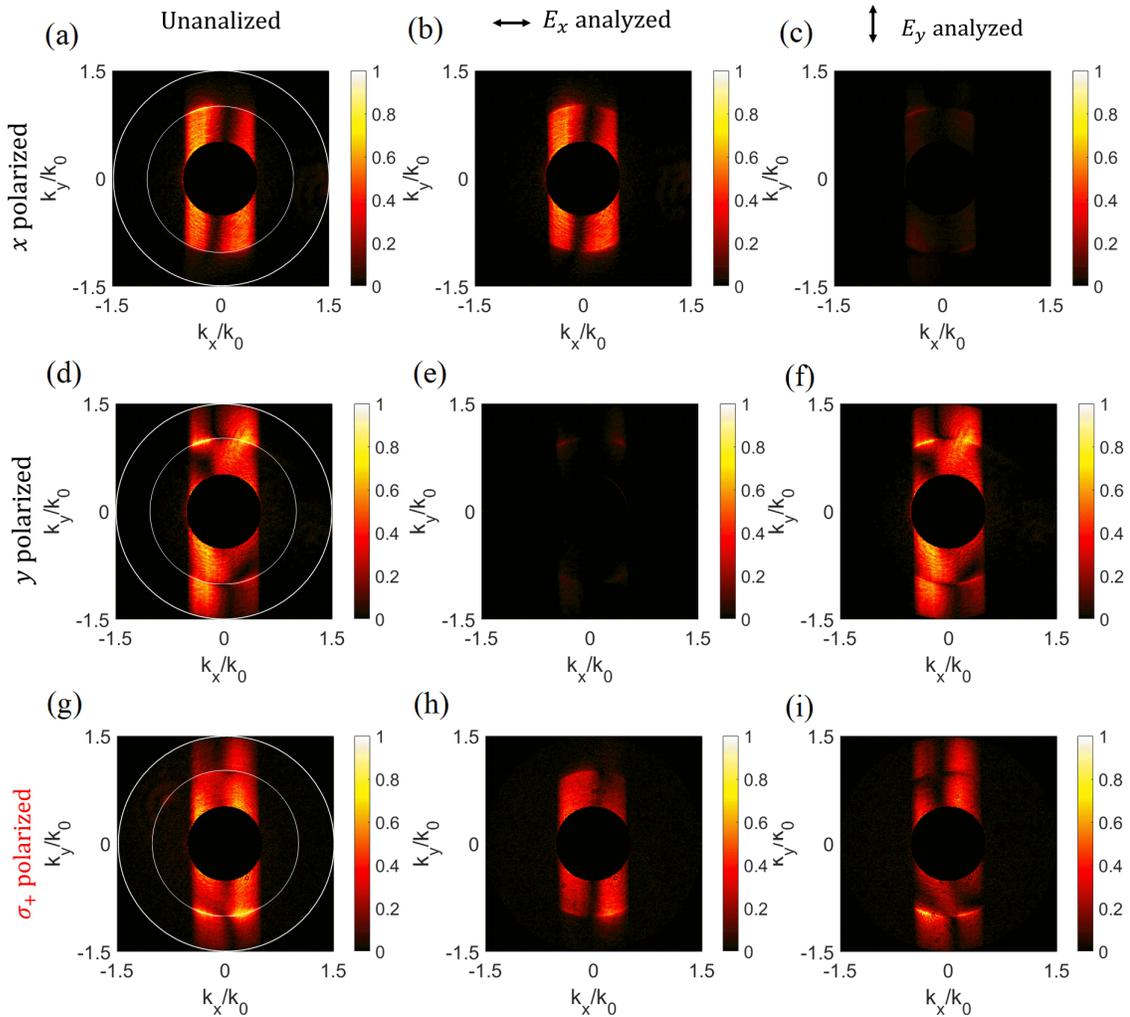

FIG. S3: Experimentally measured FP intensity distribution for $LG_0^1$ beam and their polarization analysis. (a)FP due to $x$ polarized beam. (b) and (c) analyzed along $x$ and $y$ respectively indicate almost all the scattered light is $x$ polarized. (d) FP due to $y$ polarized beam. (e) and (f) analyzed along $x$ and $y$ respectively indicate the light is predominantly $y$ polarized. (g) FP due to $\sigma_+$ LG$_0^1$ beam. (g) and (h) indicate polarization analyzed along $x$ and $y$ respectively.

## S4: NUMERICAL SIMULATION DETAILS

We have employed finite element method simulation based on commercial COMSOL 5.5 software with wave-optics module to perform numerical simulations. AgNW have been modelled using a pentagonal cross-sectional geometry of length 5 $\mu m$, with refractive index set to Ag at 633 nm wavelength, placed on a cuboid geometry set to mimic glass substrate. The LG beam is set to incident through a port of $2.5\mu m \times 2.5\mu m$ and the beam waist was kept at 633 nm to mimic the focusing conditions. Appropriate scattering boundary conditions were applied to avoid spurious reflections from the boundaries of the simulation box.

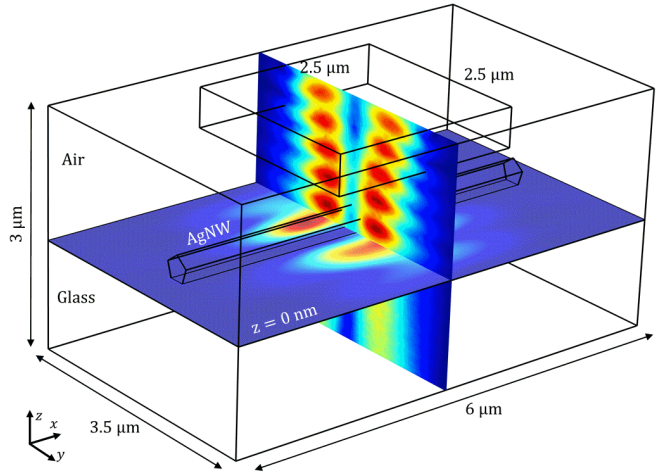

FIG. S4: The numerical simulation geometry.

## S5: $k_y < 0$ REGION ANALYSIS OF $\sigma_+$ AND $\sigma_-$ POLARIZED LG$_0^1$ BEAM

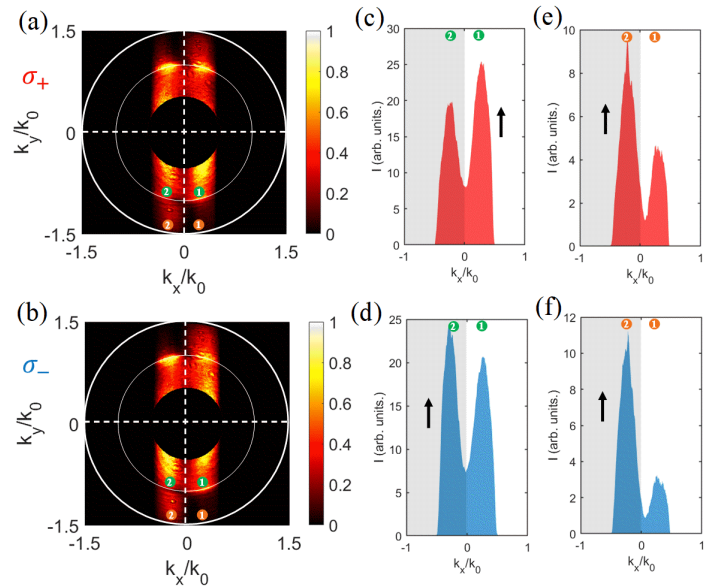

FIG. S5: Experimentally measured scattered FP intensity distribution of (a) $\sigma_+$ polarized and (b) $\sigma_-$ polarized LG$_0^1$ beam. The corresponding intensity distribution profile in the sub-critical region of the $k_y < 0$ part of FP for $\sigma_+$ and $\sigma_-$ polarized LG$_0^1$ beam is shown in (c) and (d) respectively. The same for super-critical region is shown in (e) and (f). The black arrows indicate the effective directionality. In the sub-critical region (c) $D = -0.130$ for $\sigma_+$ and (d) $D = 0.0425$ for $\sigma_-$ polarized beam. In the super-critical region (e) $D = 0.290$ for $\sigma_+$ and (f) $D = 0.526$ for $\sigma_-$ polarized beam

## S6: DIRECTIONALITY DEPENDENCE ON THE NANOWIRE SIZE

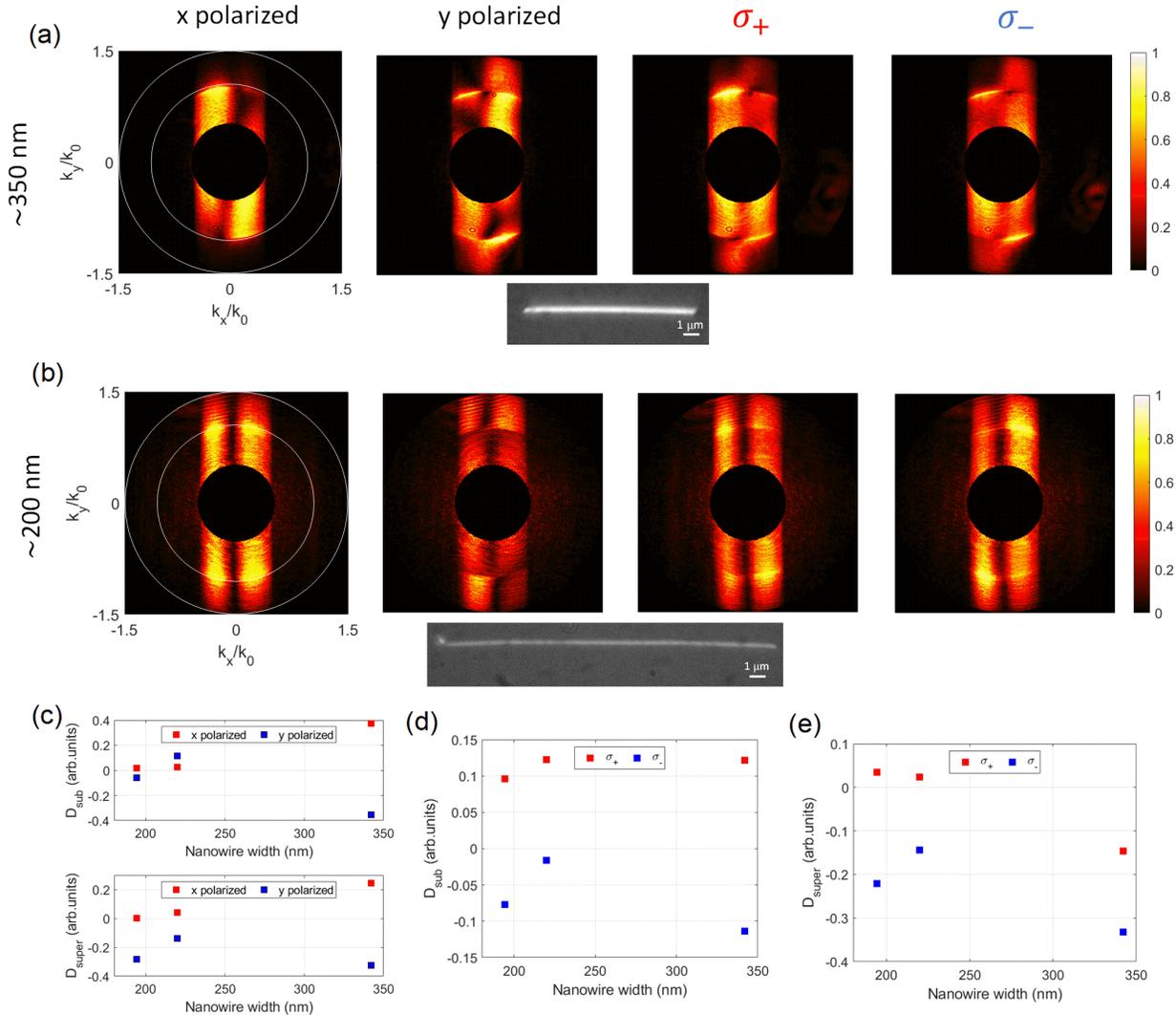

FIG. S6: Size dependence of scattering pattern for nanowire of width (a) ≈ 350 nm and (b) ≈ 200 nm. Nanowire micrographs are shown below the corresponding row. (c) Directionality values in the sub- ($D_{sub}$) and super-critical region ($D_{super}$) for $k_y > 0$ region of FP for $x$ and $y$ polarized $LG_0^1$ beam, (d) $D_{sub}$ for $\sigma_+$ and $\sigma_-$ polarized $LG_0^1$ beam. (e) $D_{super}$ for $k_y > 0$ part of FP for $\sigma_+$ and $\sigma_-$ polarized $LG_0^1$ beam.

By comparing figure S6(a) and (b) the differences in the scattering pattern due different width of the nanowire can be readily seen. For $x$ polarized $LG_0^1$ beam, scattering from ≈ 200 nm NW leads to significant signal in the super-critical region compared to ≈ 350 nm NW. As a result, the directionality values in the sub- and super-critical region in the FP for ≈ 200 wire differ from the ≈ 350 nm case, as shown in figure S6(c)-(e). For the thinner nanowires, both the magnitudes of $D_{sub}$ and $D_{super}$ for $x$ and $y$ polarized incident beam is small compared to the ≈ 350 nm wire, rendering measurement of OAM mode with ≈ 350 nm more accurate. For $\sigma_+$ or $\sigma_-$ polarized $LG_0^1$ beam, both ≈ 200 nm and ≈ 350 nm wires detect the polarization state through the inversion of the $D_{sub}$ sign, although with relatively lesser magnitude. But, for thinner nanowires $D_{super}$ may take opposite sign for $\sigma_+$ $LG_0^1$ beam, giving us incorrect reading of the OAM mode. This can be attributed to the fact that for ≈ 200 nm wires, there is significant signal in the super-critical region of FP for $x$ polarized beam.

## S7: ELLIPTICAL POLARIZATION DEPENDENCE ON THE DIRECTIONALITY

We have investigated the polarization dependence of the directionality in the sub- ($D_{sub}$) and super-critical ($D_{super}$) region of $k_y > 0$ region of the FP by changing the incident polarization through a half-wave plate (HWP) followed by a quarter-wave plate (QWP). In addition, we have also investigated the elliptical polarization dependence of the intensity in the super-critical region ($I_{super}$) of $k_y > 0$ part of the FP. The linear polarization angle (LP) due to the HWP is represented as $\alpha$ and QWP angle is represented by $\beta$, and are measured with respect to nanowire long axis ($x$ axis) as shown in figure S7(a).

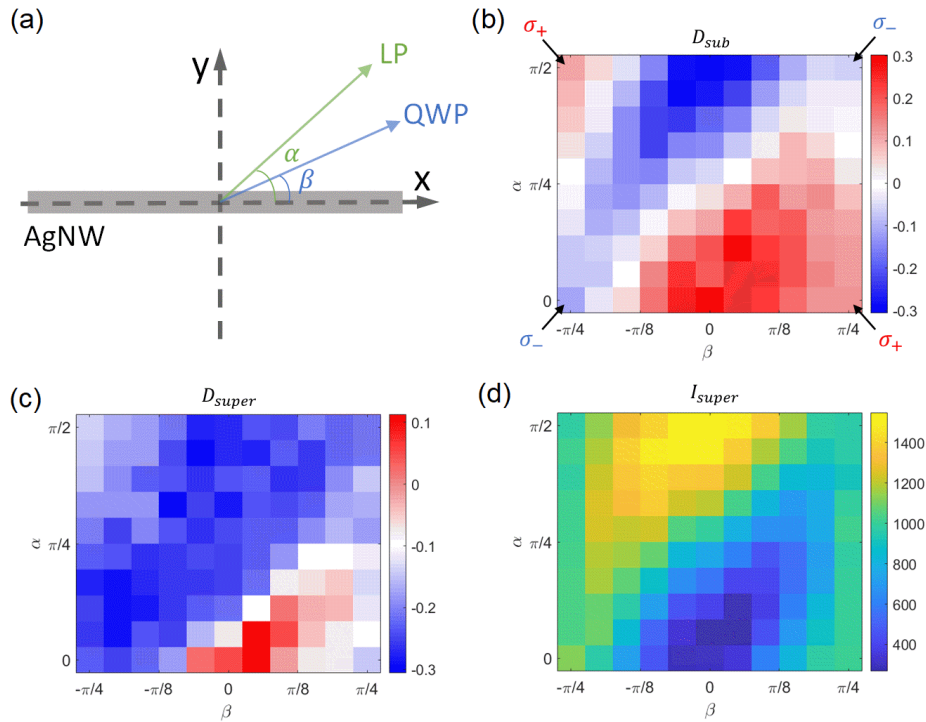

FIG. S7: Polarization dependent directionality for $LG_0^1$ beam. (a) The linear polarization (LP) angle ($\alpha$) and quarter-wave plate (QWP) angle ($\beta$) is defined with respect to nanowire long axis. Measured directionality in the (b) sub-critical region (c) super-critical region of the FP. (c) Integrated intensity in the super-critical region. The $\sigma_+$ and $\sigma_-$ states used in our experiments are indicated by black arrows and show inversion of $D_{sub}$ sign for $\sigma_+$ and $\sigma_-$ states, whereas the sign of $D_{super}$ remains the same, effectively reading out the SAM and OAM states.

Simultaneous measurement of spin and orbital angular momentum states are possible due to the reference set by the orbital energy flow, the readout of which is performed by measurement of the directionality in the super critical region of the FP. The SAM states can then be measured by comparing the directionality in the sub-critical region of the FP, which is dominantly due to the contribution of spin energy flow of circularly polarized beams.

However, it is not possible to reconstruct elliptically polarized states with the method presented here. For elliptically polarized beams, no such reference can be established based on just the directionality measurements as the directionality magnitudes in both the sub- and super-critical region depends on the polarization states as well as transverse energy flow. However, with the data presented in figure S7(b)−(d) for $D_{sub}$, $D_{super}$ and $I_{super}$ respectively, the following observations can be made:

1. The LCP ($\sigma_+$) and RCP ($\sigma_-$) state used in our experiments, indicated by black arrows in figure S7(a), show inversion of $D_{sub}$ values from $\sigma_+$ to $\sigma_-$ case, whereas $D_{super}$ remains the same, effectively reading out SAM and OAM states. The determination is performed by comparing the sign of directionality in the sub- and super-critical region.

2. To determine linear polarization states, the magnitude of the directionality also has to be considered in addition to its sign. For example, if $D_{sub} \approx -0.3$ and $D_{sub} < 0$, the state can be either determined as $y$ polarized $LG_0^1$ beam, or it can be $x$ polarized $LG_0^{-1}$ beam. To further distinguish between these two cases, the integrated intensity in the super-critical region ($I_{super}$) of the $k_y > 0$ region has to be considered. A low $I_{super}$ indicates $x$ polarized beam and high $I_{super}$ indicates $y$ polarized beam. Thus, this distinction can be made by considering the parameters: $D_{sub}$, $D_{super}$ and $I_{super}$.

3. For other cases, it is not possible to determine and reconstruct the elliptical polarization states accurately by considering the above parameters ($D_{sub}$, $D_{super}$ and $I_{super}$).

However, by studying elliptical polarization dependence of $D_{sub}$, $D_{super}$ and $I_{super}$, the elliptical polarization states can be reconstructed if either $\alpha$ or $\beta$ is known. The accuracy of the elliptical polarization measurement in such cases will be subject to standardization of the directionality values, which can vary depending on the NW width and incident beam profile.

## S8: TRANSVERSE ENERGY FLOW OF $LG_0^n$ BEAMS FROM $n = 1$ TO $n = 5$

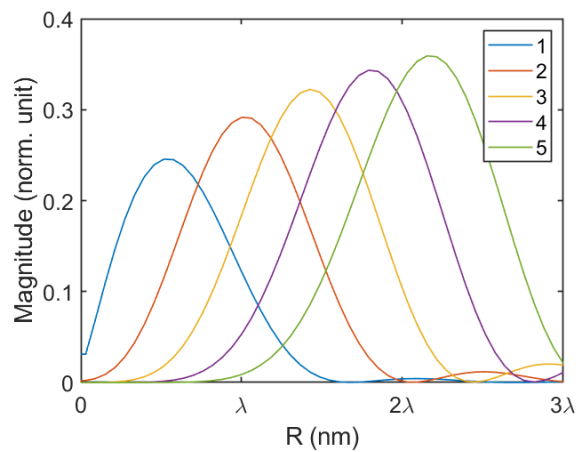

FIG. S8: Theoretically calculated normalized transverse flow density at the focal plane of 0.5 NA lens for $x$ polarized $LG_0^n$ beam from $n = 1$ to $n = 5$ exhibiting incremental OFD. The normalization is with respect to maximum of total Poynting vector magnitude.


\end{document}